\documentclass[3p,times,procedia]{elsarticle}
\usepackage{nupha_ecrc}

\volume{00}

\firstpage{1}

\journalname{Nuclear Physics A}

\runauth{}

\jid{nupha}

\jnltitlelogo{Nuclear Physics A}

\usepackage{amssymb}

\usepackage[figuresright]{rotating}

\usepackage{xcolor}

\begin{document}

\begin{frontmatter}



\dochead{XXVIIIth International Conference on Ultrarelativistic Nucleus-Nucleus Collisions\\ (Quark Matter 2019)}

\title{Anomalous magnetohydrodynamics with constant anisotropic electric conductivities}


\author[label1]{Ren-jie Wang}
\author[label2]{Patrick Copinger}
\author[label1]{Shi Pu}

\address[label1]{Department of Modern Physics, University of Science and Technology
of China, Hefei 230026, China}
\address[label2]{Department of Physics, The University of Tokyo, 
             7-3-1 Hongo, Bunkyo-ku, Tokyo 113-0033, Japan}

\begin{abstract}

We study anomalous magnetohydrodynamics in a longitudinal boost
invariant Bjorken flow with constant anisotropic electric
conductivities as outlined in Ref. \cite{Siddique:2019gqh}.
For simplicity, we consider a neutral fluid and a 
force-free magnetic field in the transverse direction.
We derived analytic solutions of the electromagnetic fields in the laboratory frame, 
the chiral density, and the energy density as functions of proper time.
\end{abstract}


\end{frontmatter}



\section{Introduction} 
Recently, several novel quantum transport phenomena related to 
strong magnetic fields have been extensively studied. 
Some examples include the chiral magnetic effect and chiral separation effect
\cite{Vilenkin1980a,Kharzeev:2007jp,Fukushima:2008xe}.
Similarly, electric fields
 can also induce the chiral separation
\cite{Huang:2013iia,Pu:2014cwa,Jiang:2014ura,Pu:2014fva}
and chiral Hall separation effects\cite{Pu:2014fva}. 
There are also other
non-linear chiral transport effects, 
such as are discussed in Refs.
\cite{Chen:2016xtg,Chen:2013iga,Ebihara:2017suq,Hidaka:2017auj} and 
the connection 
to Schwinger 
pair production \cite{Copinger:2018ftr} 
(also see the reference therein). 

Anomalous magnetohydrodynamics (anomalous MHD), which is 
the relativistic magnetohydrodynamics in the presence of the chiral magnetic effect
and chiral anomaly, is a framework in which to study the aforementioned chiral transport phenomena.
In Refs. \cite{Pu:2016ayh,Roy:2015kma} we derived
analytic solutions for 
ideal MHD with longitudinal boost
invariance and a transverse magnetic field. 
In the latter part of Ref. \cite{Pu:2016ayh}, we considered magnetization effects.
 Following this framework, the studies have been extended to 
2+1 dimensional MHD with 
Bjorken flow \cite{Pu:2016bxy,Pu:2016rdq} and Gubser flow \cite{Shokri:2018qcu}.
The numerical simulation of ideal MHD can be found 
in Ref. \cite{Inghirami:2016iru}.

Very recently, we obtained analytic
solutions for anomalous MHD with
Bjorken flow \cite{Siddique:2019gqh}. In that work, 
we only consider anomalous MHD with 
a constant electric conductivity, $\sigma$.
In a strong magnetic field, the
electric conducting flow will be $\bold{j}^i=\sigma^{ij} \bold{E}^j$, where
$\sigma^{ij}$ is the anisotropic electric conductivity tensor.
The tensor  $\sigma^{ij}$
includes the classical Hall conductivity, $\sigma_H$, and
the electric conductivities parallel and perpendicular
to the magnetic fields, i.e., $\sigma_{\parallel}$ and $\sigma_{\bot}$, respectively.  
$\sigma_{\parallel}$ and $\sigma_{\bot}$ have been
computed by a
sum over all Landau levels \cite{Fukushima:2017lvb}. 
Since the magnetic fields are extremely strong in relativistic heavy ion collisions,  
we also need to consider anomalous
MHD with anisotropic electric conductivities.

In Ref. \cite{Fukushima:2017lvb}, 
the authors
have found that both $\sigma_{\parallel}$ and $\sigma_{\bot}$ depend
on the temperature and magnetic field strength.
For simplicity, in the present work, we assume that
$\sigma_H,\sigma_\parallel$, and $\sigma_\bot$ are all constants.
We will present the 
results for a temperature and magnetic field dependent $\sigma^{ij}$ somewhere else. 


The structure of this work is as follows. In Sec. \ref{sec:MHD-with-CME}, we derive 
an analytic solution for
anomalous MHD with constant anisotropic electric conductivities. 
We then summarize in Sec. \ref{sec:summary}.
Throughout
this work, we will use the metric $g_{\mu\nu}=\mathrm{diag}\{+,-,-,-\}$,
and choose Levi-Civita tensor satisfying $\epsilon^{0123}=-\epsilon_{0123}=+1$.

\section{Anomalous magnetohydrodynamics with constant anisotropic electric conductivities}
\label{sec:MHD-with-CME}

The main equations for MHD 
are the energy-momentum and currents
conservation equations coupled with
Maxwell's equations.
(See 
e.g., Ref. \cite{Pu:2016ayh,Roy:2015kma,Pu:2016bxy,Pu:2016rdq,Huang:2009ue}
 and references therein
for details). Here, we will neglect other dissipative effects, 
such as the
bulk viscous pressure, shear viscous tensor,
and heat conducting flow. 
We follow the framework and 
assumptions as used
in our previous work  \cite{Siddique:2019gqh}.
The energy-momentum conservation equation without viscous effects reads 
\begin{equation}
\partial_{\mu}T^{\mu\nu}=0\,,\label{eq:EMT_01}
\end{equation}
where $T^{\mu\nu}$ is the full energy momentum tensor and can
be decomposed as,
\begin{equation}
T^{\mu\nu}=(\varepsilon+p+E^{2}+B^{2})u^{\mu}u^{\nu}-(p+\frac{1}{2}E^{2}+\frac{1}{2}B^{2})g^{\mu\nu}\nonumber  -E^{\mu}E^{\nu}-B^{\mu}B^{\nu}-u^{\mu}\epsilon^{\nu\lambda\alpha\beta}E_{\lambda}B_{\alpha}u_{\beta}-u^{\nu}\epsilon^{\mu\lambda\alpha\beta}E_{\lambda}B_{\alpha}u_{\beta}\,,\label{eq:totEMtensor_01-1}
\end{equation}
where $\varepsilon$ and $p$ are energy density and pressure, respectively. 
Here, we have introduced the four-vector form of electromagnetic fields in a co-moving frame of a fluid cell, 
\begin{equation}
E^{\mu}=F^{\mu\nu}u_{\nu}\,, \qquad B^{\mu}=\frac{1}{2}\epsilon^{\mu\nu\alpha\beta}u_{\nu}F_{\alpha\beta}\,, \qquad
E^{\mu}E_{\mu} \equiv -E^{2}\,, \qquad B^{\mu}B_{\mu} \equiv-B^{2}\,.
\label{eq:EB_def01}
\end{equation}
The conservation equations 
for the currents are 
\begin{equation}
\partial_{\mu}j_{e}^{\mu} = 0\,,  \qquad
\partial_{\mu}j_{5}^{\mu} = -e^{2}CE\cdot B\,,\label{eq:currents_con_01}
\end{equation}
where $j_{e}^{\mu}$ and $j_{5}^{\mu}$ are the electric
and chiral (axial) current, respectively,
and $C=1/(2\pi^{2})$ refers to the
chiral anomaly. These currents can be decomposed as
\begin{equation}
j_{e}^{\mu} = n_{e}u^{\mu}+\sigma^{\mu\nu}E_{\nu}+\xi B^{\mu}\,,  \qquad
j_{5}^{\mu} = n_{5}u^{\mu}+\sigma_{5}E^{\mu}+\xi_{5}B^{\mu}\,,\label{eq:current_02}
\end{equation}
where $n_{e}$ and $n_{5}$ are the electric and chiral charge densities,
respectively, and $\sigma_{5}$ is the chiral
electric conductivity \cite{Huang:2013iia,Pu:2014cwa,Pu:2014fva}.
The coefficients $\xi=eC\mu_{5}$ and $\xi_{5}=eC\mu_{e} $ are associated with the chiral magnetic and separation effects 
\cite{Fukushima:2008xe,Gao:2012ix,Chen:2012ca},
with $\mu_e$ and $\mu_5$ being the electric and chiral chemical potential, respectively.
The anisotropic electric conductivity tensor can be written as \cite{Fukushima:2017lvb}, 
\begin{equation}
\sigma^{\mu\nu}=\sigma_{H}\epsilon^{\mu\nu\alpha\beta}u_{\alpha}\frac{B_{\beta}}{B}-\sigma_\parallel \frac{B^{\mu}B^{\nu}}{B^{2}}+\sigma_{\bot}\left(g^{\mu\nu}+\frac{B^{\mu}B^{\nu}}{B^{2}}\right)\,, \label{eq:conductivity}
\end{equation}
where $\sigma_{H},\sigma_\parallel,$ and $\sigma_{\bot}$ 
denote the classical Hall, longitudinal,
and transverse conductivity, respectively.
We use the covariant form for Maxwell's equations,
\begin{equation}
\partial_{\mu}F^{\mu\nu}  =  j_{e}^{\nu}\,,  \qquad
\partial_{\mu}(\epsilon^{\mu\nu\alpha\beta}F_{\alpha\beta})  =  0\,.\label{eq:Maxwell_01b}
\end{equation}
We also choose the equation of state in a hot fireball limit,
\begin{equation}
\varepsilon =  c_{s}^{-2}p\,, \qquad n_{e}  =  a\mu_{e}T^{2}\,, \qquad n_{5}  =  a\mu_{5}T^{2}\,,\label{eq:eos_02}
\end{equation}
where $a$ is again a dimensionless constant and $T$ is the temperature.
For an ideal fluid,
we have $a=1/3$ \cite{Gao:2012ix, Pu:2011vr}.


In our previous work \cite{Siddique:2019gqh}, we derived
a self-consistent analytic
solution for anomalous MHD in a Bjorken flow.  For simplicity, 
we assume the fluid is charge neutral, 
i.e., 
$\mu_e=n_e=0$. 
This is applicable since the chiral electric
conductivity, $\sigma_{5}$, is proportional to
$\mu_e$ as  $\sigma_{5}\propto\mu_{e}\mu_{5}$
in the small $\mu_{e}$ and $\mu_{5}$ 
limits
\cite{Huang:2013iia,Pu:2014cwa,Pu:2014fva}. Therefore,
in this case, $\sigma_5 \simeq 0$. Similarly, $\xi_5\propto \mu_e$ also vanishes. 
We also assume that the system is 
longitudinally
boost invariant and 
that the
electromagnetic fields in the longitudinal direction are negligible. 
The fluid velocity in a Bjorken flow reads, 
\begin{equation}
u^{\mu}=\left(\cosh\eta,0,0,\sinh\eta\right)=\gamma(1,0,0,z/t),\label{eq:Bjokren_velocity_01}
\end{equation}
with $\tau=\sqrt{t^2-z^2}$
 and $\eta=\frac{1}{2}\ln[(t+z)/(t-z)]$ being  the proper time and
the space-time rapidity, respectively.

Usually, the electromagnetic fields can accelerate the fluid velocity through the Lorentz force. 
In our case, 
however,
we have found a special 
field configuration fields that keeps
the fluid force-free,
\begin{equation}
E^{\mu}=(0,0,\chi E(\tau),0),\;B^{\mu}=(0,0,B(\tau),0),\label{eq:EB_02}
\end{equation}
where 
$\chi=\pm1$,
and without loss of generality, we only consider 
fields in the $y$ direction.
From Eq. (\ref{eq:EMT_01}), we have checked that  $(g_{\mu\nu}-u_\mu u_\nu) (\partial_{\rho}T_{Matter}^{\rho\nu}-J_{\rho}F^{\rho\nu})=0$,
where $T_{Matter}^{\rho\nu}=\left.T^{\rho\nu}\right|_{E,B\rightarrow0}$,
 is automatically satisfied 
according
our assumptions, 
i.e., the electromagnetic fields will not accelerate the 
fluid.

Our main equations are  $u_\nu \partial_\mu T^{\mu\nu}=0$ from Eq. (\ref{eq:EMT_01}) ,
coupled with Maxwell's equations, (\ref{eq:Maxwell_01b}), and their corresponding 
constitution equations, (\ref{eq:totEMtensor_01-1}, \ref{eq:EB_def01}, \ref{eq:current_02}, \ref{eq:conductivity}, \ref{eq:EB_02}). After some calculations, those coupled equations reduce
to
\begin{eqnarray}
\frac{d}{d\tau}E+\frac{1}{\tau}E+\sigma_\parallel E+\chi\xi B=0\,, &  & \frac{d}{d\tau}B+\frac{B}{\tau}=0\,,  \label{eq:Maxwell_02a} \\
\frac{d}{d\tau}\varepsilon+(\varepsilon+p)\frac{1}{\tau}-\sigma_\parallel E^{2}-\chi\xi EB=0\,, & & 
\frac{d}{d\tau}n_{5}+\frac{n_{5}}{\tau} =  e^{2}C\chi EB\,. \label{eq:energy_density_03}
\end{eqnarray}
We notice that Eqs. (\ref{eq:energy_density_03}) are similar to the 
Eqs.
(23, 24, 25, 27) in Ref. \cite{Siddique:2019gqh} by
replacing $\sigma$ 
with $\sigma_\parallel$, a reasonable replacement:
For the electromagnetic field configuration
in Eq.(\ref{eq:EB_02}),
the electric field is parallel to the magnetic field. Therefore, 
from Eq. (\ref{eq:conductivity}), only $\sigma_\parallel$ will
contribute to our 
final result.

Using
the  non-conserved
charges method \citep{Csorgo:2003rt,Shokri:2017xxn},
we obtain 
analytic
solutions for Eqs. (\ref{eq:energy_density_03}) 
with the 
EoS, Eq. (\ref{eq:eos_02}),
in a small $\mu_5/T$ limit, 
\begin{eqnarray}
E(\tau) & = & E_{0}\left(\frac{\tau_{0}}{\tau}\right)\left\{ e^{-\sigma_\parallel(\tau-\tau_{0})}-a_{1}e^{-\sigma_\parallel\tau}[\textrm{E}_{1-2c_{s}^{2}}(-\sigma_\parallel \tau_{0})-\left(\frac{\tau}{\tau_{0}}\right)^{2c_{s}^{2}}\textrm{E}_{1-2c_{s}^{2}}(-\sigma_\parallel \tau)]+\mathcal{O}(a_{i}^{2})\right\} \,,\nonumber \\
n_{5}(\tau) & = & n_{5,0}\left(\frac{\tau_{0}}{\tau}\right)\left\{ 1+a_{2}e^{\sigma_\parallel \tau_{0}}[\textrm{E}_{1}(\sigma_\parallel\tau_{0})-\textrm{E}_{1}(\sigma_\parallel \tau)]+\mathcal{O}(a_{i}^{2})\right\} \,,\nonumber \\
\varepsilon(\tau) & = & \epsilon_{0}\left(\frac{\tau_{0}}{\tau}\right)^{1+c_{s}^{2}}\left\{ 1+\sigma_{\parallel}\frac{E_{0}^{2}}{\varepsilon_{0}}e^{2\sigma_{\parallel}\tau_{0}}[\tau_{0}\textrm{E}_{1-c_{s}^{2}}(2\sigma_{\parallel}\tau_{0})-\tau\left(\frac{\tau}{\tau_{0}}\right)^{c_{s}^{2}-1}\textrm{E}_{1-c_{s}^{2}}(2\sigma_{\parallel}\tau^{\prime})]\right.\nonumber \\
 &  & \left.+\frac{a_{3}}{\tau_{0}}e^{\sigma_\parallel \tau_{0}}[\tau_{0}\textrm{E}_{2-3c_{s}^{2}}(\sigma_\parallel \tau_{0})-\tau\left(\frac{\tau_{0}}{\tau}\right)^{2-3c_{s}^{2}}\textrm{E}_{2-3c_{s}^{2}}(\sigma_\parallel \tau)]+\mathcal{O}(a_{i}^{2},a_{i}E_{0}^{2}/\varepsilon_{0})\right\} \,.\label{eq:sol_02}
\end{eqnarray}
where $\textrm{E}_{n}(z)\equiv\int_{1}^{\infty}dtt^{-n}e^{-zt}$ is
the generated exponential integral. The coefficients $a_{1}$, $a_{2}$ and $a_{3}$
are dimensionless constants determined by the initial conditions,
\begin{equation}
a_{1}  =  eC\chi\frac{B_{0}n_{5,0}}{aT_{0}^{2}E_{0}}\tau_{0}\,, \qquad
a_{2}  =  \frac{e^{2}C\chi E_{0}B_{0}}{n_{5,0}}\tau_{0}\,, \qquad
a_{3}  =  \frac{eC\chi}{a}\frac{n_{5,0}E_{0}B_{0}}{\varepsilon_{0}T_{0}^{2}}\tau_{0}\,,
\end{equation}
where the lower index $0$ denotes the quantity at the initial proper 
time, $\tau_0$.

The electromagnetic fields in the laboratory frame 
are
given by, 
\begin{equation}
\mathbf{E}_{\textrm{lab}} = (\gamma v^{z}B(\tau),\;\chi\gamma E(\tau),\;0)\,, \qquad
\mathbf{B}_{\textrm{lab}} = (-\gamma v^{z}\chi E(\tau),\;\gamma B(\tau),\;0)\,,\label{eq:EB_lab}
\end{equation}
where $B(\tau)=B_0 (\tau_0 / \tau)$, 
$\chi=\pm1$,
and $E(\tau)$ is given by 
Eq. (\ref{eq:sol_02}). 
We find 
that
$\bold{B}^y_{\textrm{lab}} \propto 1 / \tau$ and $\bold{B}^x_{\textrm{lab}} \propto \exp(-\sigma_\parallel \tau) / \tau$,
i.e., the magnetic field decays much slower in this case than in the vacuum \cite{Kharzeev:2007jp}.

\section{Summary}\label{sec:summary}
\label{sec:Summary-and-conclusion}
In this work, we have studied 
anomalous
MHD with anisotropic electric conductivities. 
The fluid expands along the longitudinal direction 
with the
Bjorken boost invariant. 
For simplicity, we consider 
a
charge neutral fluid and a particular 
force-free electromagnetic field configuration
, i.e., 
one that does not accelerate the fluid velocity. 
We have derived 
an
analytic solution for anomalous MHD in our cases in a small $\mu_5$ limit. 
In the future, we plan to extend the discussion 
to
cases with temperature and magnetic field dependent 
anisotropic electric conductivities.





\bibliographystyle{elsarticle-num}
\bibliography{MHD}







\end{document}